\documentclass[12pt]{article}
\usepackage{amssymb, amsmath, xspace,% showkeys
}
\renewcommand{\a}{\alpha}
\renewcommand{\b}{\beta}
\newcommand{\nad}[2]{\genfrac{}{}{0pt}{}{#1}{#2}}

\newcommand{\p}{\partial}

\newcommand{\C}{\mathbb C}

\newcommand{\Z}{\mathbb Z}

\newcommand{\vf}{\varphi}
\newcommand{\cl}{{\cal L}}
\newtheorem{theorem}{Theorem}

\def\beq#1#2\eeq{%
           \begin{equation}%
           \label{#1}%
               #2%
           \end{equation}%
       }

\begin{document}

\begin{center}

{\bf\Large Quasiinvariants of Coxeter groups and \pmb{$m$}-harmonic
polynomials}

\vspace{1cm}

{\bf  M. Feigin$^1$ and A. P. Veselov$^{1,2}$}
\end{center}

\vspace{1cm}

\noindent $^1$
Department of Mathematical Sciences,
Loughborough University,\\
Loughborough,  LE11 3TU, UK

\noindent $^2$ Landau Institute for Theoretical Physics,
Kosygina 2, Moscow, 117940, Russia

\noindent E-mail addresses: M.Feigin@lboro.ac.uk,
A.P.Veselov@lboro.ac.uk

\vspace{1cm}

\begin{abstract}
\noindent
The space of $m$-harmonic polynomials related to a Coxeter group $G$ and a multiplicity function
$m$ on its root system is defined as the joint kernel of the properly gauged invariant integrals of the
corresponding generalised quantum Calogero-Moser problem.
The relation between this space and the ring of all quantum integrals of this system (which is
isomorphic to the ring of corresponding quasiinvariants) is investigated.
\end{abstract}

\vspace{1cm}

\section{Introduction}

Let $G$ be any Coxeter group, i.e. a finite group generated by
reflections with respect to some hyperplanes in a Euclidean space
$V$ of dimension $n$. Let $\Sigma$ be a set of the hyperplanes
$\Pi_\a:(\a,x)=0$
corresponding to all the reflections $s_\a\in G$ and let $A$ be a
set of the corresponding (arbitrarily chosen) normals $\a$.
Let us also consider a $G$-invariant ${\Z}_{\ge 0}$ -valued function $m$ on
$\Sigma$ which will be called {\it multiplicity}.
 In other words to any hyperplane $\Pi_\a\in\Sigma$
we prescribe a nonnegative integer $m_\a$ such that if $\Pi_\a=g(\Pi_\b),\,
g\in G$ then $m_\a=m_\b$.

Let $S = S(V)$ be the ring of all polynomials on $V$, $S^G$ be the subring of
$G$-invariant polynomials. According to classical Chevalley result \cite{Chev}
$S^G$ is freely generated by some homogeneous polynomials
$\sigma_1,\ldots,\sigma_n$.

The main object of our investigation is the following subring
$Q_m=Q_m(\Sigma)\subset S(V)$. It consists of
the polynomials $q$ which are invariant up to order $2 m_\a$
 with respect to any reflection $s_\a$:
\beq{quasi}
q(s_\a(x))=q(x) + o\left((\a,x)^{2m_\a}\right)
\eeq
near the hyperplane $(\a,x)=0$ for any $\a\in A$.
Equivalently, for any $\a\in A$ the normal derivatives $\p_\a^s q = (\a,\frac{\p}{\p x})^s q$
 must vanish on $\Pi_\a$ for $s=1,3,5,\ldots,
2m_\a-1$:
$$
\p_\a^s q|_{\Pi_\a}=0.
$$
We will call these polynomials {\it
$m$--quasiinvariants} of the Coxeter group $G$ or simply {\it quasiinvariants}.

The rings $Q_m$ have been introduced in the theory of quantum
Calogero--Moser systems by O.Chalykh and one of the authors \cite{CV}.
It has been shown \cite{CV,VSCh} that for any Coxeter group $G$ and any integer-valued
multiplicity function $m$
 there exists a homomorphism
$$
\varphi_m: Q_m\to D_{\Sigma}(V),
$$
where $D_{\Sigma}(V)$ is the ring of all differential operators in $V$
with rational coefficients from the algebra generated by $(\alpha, x)^{-1}, \alpha \in A$ and constant functions (see the next section for the details).
 In particular, for $q=x^2$ (which is obviously invariant and therefore quasiinvariant)
 the
corresponding operator $\varphi_m(q)$ is the generalised
Calogero--Moser operator

\beq{CM}
L=\Delta-\sum_{\a\in A}\frac{m_\a(m_\a+1)(\a,\a)}{(\a,x)^2}
\eeq
first introduced by Olshanetsky and Perelomov \cite{OP}.
 It will be more convenient for us to use the gauge
transformation ${\cal L}= \hat g L \hat g^{-1}$, where
$\hat g$ is the operator of multiplication by
$g=\prod(\a,x)^{m_\a}$, after which the operator $L$ takes the form
\beq{lcur}
{\cal L} = \Delta -\sum_{\a\in A}\frac{2m_\a}{(\a,x)}\p_\a.
\eeq
This gauge is natural from the point of view of the theory of
symmetric spaces where such operators appear as the radial parts of
the Laplace--Beltrami operators (see e.g.~\cite{Helg}).
Let
$$
\chi_m: Q_m \to D_{\Sigma}(V)
$$
be the corresponding
gauged version of the homomorphism $\varphi_m$:
$$
\chi_m(q) = \hat g \varphi_m(q) \hat g^{-1}.
$$

%and  $${\cal L}_i=\chi_m(\sigma_i)=\hat g L_i \hat g^{-1}, i =1,...,n$$
%be the gauged Calogero-Moser integrals $L_1, ..., L_n$.

We should mention that for a generic (not necessary integer-valued) multiplicity
function there exists an isomorphism (sometimes called as Harish-Chandra
isomorphism, see
\cite{HO,H}) between $S^G$ and the ring $D^G_m$ of $G$-invariant quantum integrals
of the Calogero-Moser problem (\ref{CM}):
$$\gamma_m: S^G \cong D^G_m,$$
where $$D^G_m = \{{\cal D} \in D_{\Sigma}(V): [{\cal L}, {\cal D}] = 0, g({\cal D})={\cal D} \,\, \mbox{for all}\,\, g \in G\}.$$
As a corollary we have usual integrability for (\ref{lcur}) with $n$
commuting quantum integrals
${\cal L}_1={\cal L}, {\cal L}_2,...,{\cal L}_n$ corresponding to some basic invariants
$\sigma_1 = x^2, \sigma_2,...,\sigma_n$.

An important novelty of \cite{CV} was the possibility of the extension
of $\gamma_m$
to a much bigger ring $Q_m$ in the case when all $m_\a$ are integer,
which implies the algebraic integrability of the corresponding quantum
Calogero-Moser problem (see \cite{VSCh} for details).

The primary goal of the present paper is to explain that all the information
about the additional quantum integrals of Calogero-Moser problem is
actually contained in the joint kernel of its standard invariant integrals ${\cal L}_1,...,{\cal L}_n.$
More precisely,
let us define the space $H_m$ as the solutions of the
following system
\begin{equation}
\label{systfirst}
\begin{cases}
\cl_1\psi=0\\
\hbox to 15mm {\dotfill}\relax\\
\cl_n\psi=0
\end{cases}
\end{equation}
We will show that all the solutions of
(\ref{systfirst}) are polynomial and that all these polynomials are $m$-quasiinvariant.
We will call these polynomials {\it m-harmonic}.
In the case $m=0$ we have the space of the usual harmonic polynomials related to
Coxeter group $G$ (see e.g \cite{Helg}).

The following linear map $\pi_m$ from $Q_m$ to $H_m$ will play the central
role in our considerations. Let us introduce $m$-discriminant
$$
w_m=\prod_{\a\in A} (\a,x)^{2m_\a+1}
$$
which obviously is $m$-quasiinvariant. We will show that $w_m$ is also
$m$--harmonic. Let $q$ be any
quasiinvariant, ${\cal L}_q=\chi_m(q)$ be the corresponding
differential operator. The map $\pi_m$ is defined by the formula
$$
\pi_m(q)={\cal L}_q (w_m).
$$
Since ${\cal L}_q$ commutes with ${\cal L}_i$ it preserves the
space $H_m$, so $\pi_m(q)\in H_m$. The question now is what is the
kernel of $\pi_m$. It is easy to show that the kernel of $\pi_m$ contains the ideal
$I_m\subset Q_m$ generated by the invariants
$\sigma_1,\ldots,\sigma_n$.

We conjecture that the following statements are true for any
Coxeter group $G$ and multiplicity function $m$.
\\

\noindent
{\bf Conjecture 1}
{\it Kernel of $\pi_m$ coincides with the ideal $I_m$.}\\

\noindent
Consider the restriction of the map $\pi_m$ onto the subspace
$H_m\subset Q_m$.
\\

\noindent
{\bf Conjecture 2}
{\it The linear map
$$
\pi_m |_{H_m}: H_m \to H_m
$$
is an isomorphism.}\\

\noindent
As a corollary we have the following isomorphism
$$
Q_m/I_m \cong H_m,
$$
and the fact that $Q_m$ is generated by $H_m$ over $S^G$.
\\

\noindent
{\bf Conjecture 3} {\it The ring $Q_m$ is a free module over $S^G$
generated by any basis in $H_m$. }\\

\noindent
In algebraic terminology this implies that $Q_m$ is a Cohen-Macauley
ring (see e.g. \cite{Benson}).
We believe that $Q_m$ is actually a Gorenstein ring.
This follows from the following Conjecture $2^*$ which can be considered
as a stronger version of Conjecture~2.

Let us introduce the following bilinear form on the space $H_m$: 
$$
<p,q>=({\cal L}_p {\cal L}_q w_m)(0),
$$
where ${\cal L}_p=\chi_m(p), {\cal
L}_q=\chi_m(q)$.
\\

\noindent
{\bf Conjecture \pmb{$2^*$}} {\it The form $<,>$ on $H_m$ is non-degenerate.}
\\

\noindent
This implies Conjecture 2. Indeed, if $q\in Ker\pi_m|_{H_m}$ then by
definition ${\cal L}_q(w_m)=0$ and therefore
$<q,p>=0$ for all $p\in H_m$.
Conjectute $2^*$ also implies that the dimensions $h_k$ of the spaces
of $m$-harmonics of degree $k$ satisfy
the duality relation
$$
h_k = h_{{\cal N}-k},
$$
where ${\cal N}=\sum(2m_\alpha+1)$ is the degree of $w_m$.

In this paper we prove all these conjectures for all the
Coxeter groups of rank 2 (i.e. for the dihedral groups $I_2(N)$) under
additional assumption that all the multiplicities are equal.
In this case we describe all the m-harmonic polynomials explicitly.

As a corollary we show that the Poincare series for the
quasiinvariants
of dihedral group $I_2(N)$ is given by
$$
p(Q_m^{I_2(N)},t)=\frac{1+2t^{(mN+1)}+\ldots+2t^{(mN+N-1)}+t^{(2m+1)N}}{(1-t^{2})(1-t^{N})}.
$$

Notice that Conjecture 3 and Chevalley theorem imply that the Poincare series of the quasiinvariants of
any Coxeter group has the form
$$
p(Q_m,t)=\frac{P(H_m,t)}{\prod_{i=1}^n (1-t^{d_i})},
$$
where $P(H_m,t)$ is the corresponding Poincare polynomial of the $m$-harmonics:
$$
P(H_m,t) = \sum_{k=0}^{\cal N} h_k t^k.
$$

Some formulas for the polynomials  $P(H_m,t)$ have been recently found in \cite{FV}
(see Concluding remarks).

\section{Quasiinvariants and quantum integrals\\ of Calogero--Moser
systems}

Let us first discuss the homomorphism
$\chi_m$ in more details. We will need some facts from the
theory of multidimensional Baker--Akhiezer functions
related to a Coxeter configuration of hyperplanes (see
\cite{CV}, \cite{VSCh}, \cite{CFV}). Let us remind that
we are using the gauge which is different from the chosen
in these papers by $g(x)=\prod(\a,x)^{m_\a}$.

For any
Coxeter group $G$ and multiplicity function $m$ there exists a
Baker--Akhiezer function (BA function) of the form
\beq{BA}
\psi=P(k,x)e^{(k,x)}
\eeq
with the following properties:
\begin{itemize}
\item
$P(k,x)$ is a polynomial in $k \in V$ and $x \in V$ with the highest term
$$
g(k)g(x)=\prod_{\a\in A} (\a,k)^{m_\a} (\a,x)^{m_\a}.
$$
\item
$\psi$ satisfies the quasiinvariance conditions in $k$-space
$$
\psi(s_\a(k))-\psi(k)=o((\a,k)^{2 m_\a}) \ \mbox{near} \ (\a,k)=0.
$$
\end{itemize}
It is known that such function does exist, and is unique and
symmetric with respect to $x$ and $k$:
$$
\psi(k,x)=\psi(x,k)
$$
(see \cite{CV}, \cite{VSCh}).
As it has been explained in \cite{CV} for any quasiinvariant $q\in
Q_m$ there exists a differential operator $\chi_m(q)={\cal L}_q(x,\frac{\p}{\p
x})$ such that
$$
{\cal L}_q(x,\frac{\p}{\p x})\psi(x,k)=q(k)\psi(x,k).
$$
The procedure of finding ${\cal L}_q$ is effective provided the
formula for $\psi$ is given. Since for $q=k^2$ we have the (gauged)
Calogero-Moser operator (\ref{lcur}) we have the following

\begin{theorem}{\rm \cite{CV,VSCh}}
For any Coxeter group $G$ and integer-valued multiplicity $m$ there exists a
homomorphism $\chi_m: Q_m \to D_{\Sigma}(V)$ mapping the algebra of
quasiinvariants $Q_m$ into the commutative algebra of quantum
integrals of generalised Calogero--Moser problem.
\end{theorem}

One can write down the following explicit formula for this homomorphism
first suggested by Yu.Berest \cite{Ber}:
\begin{equation}
\label{chi}
\chi_m(q)=c (ad_{\cal L})^{d(q)}\hat q.
\end{equation}
Here ${\cal L}$ is the gauged Calogero-Moser operator (\ref{lcur}),
 $ad_L A = L A - A L$, $\hat q$ is the
operator of multiplication by $q$, $d(q)$ is degree of
polynomial $q$ and the constant $c(q)=(2^{d(q)} d(q)!)^{-1}$.

Indeed because of the symmetry of $\psi$ with respect to $x$ and $k$ we have
$$
{\cal L}(k,\frac{\p}{\p k})\psi(x,k)=x^2\psi(x,k).
$$
Thus $\psi$ satisfies the so-called bispectral problem in the sense of
Duistermaat and Gr\"unbaum and one can use the general identity
(1.8) from their paper \cite{DG} which states that
$$
(ad_{\cal L})^r (q) [\psi] = (ad_{\hat x^2})^r({\cal L}_q)[\psi].
$$
For  $r=\deg q$ we arrive at the formula (\ref{chi}).

As we have mentioned in the Introduction the restriction $\chi_m$ onto the subring of invariants $S^G$
gives an isomorphism $\gamma_m$ between $S^G$ and the ring $D^G_m$ of invariant integrals
of the Calogero-Moser quantum problem in the gauge (\ref{lcur}). The following result
shows that the map $\chi_m$ defined in the Theorem 1 is in a certain sense
the maximal extension of this map.

Let $D_m$ be the maximal commutative ring
of differential operators on $V$  with rational coefficients which contains $D^G_m$ as a subring.

\begin{theorem}
The map $\chi_m$ is an isomorphism between the ring of $m$-quasiinvariants $Q_m$
and the ring $D_m$.
\end{theorem}

\noindent
The proof follows from the following lemma.
Let $\sigma_1=k^2, \sigma_2,..., \sigma_n$ be some generators of
$S^G$, and let
 ${\cal L}_1=\chi_m(\sigma_1) = {\cal L},\ldots,{\cal
L}_n=\chi_m(\sigma_n)$ be the corresponding invariant integrals of the
Calogero-Moser problem.\\
%\vspace{1mm}

\noindent
{\bf Lemma 1}
{\em Let} $A$ {\em be a differential operator commuting with all} ${\cal L}_i, i=1,...,n.$
{\em Then} $A = {\cal L}_q = \chi_m(q)$ {\em for some quasiinvariant} $q \in Q_m$.
\\

\noindent
To prove this let us notice that since $A$ commutes with all ${\cal L}_i$
it preserves their joint eigenspace  $V(k)$ consisting of the solutions of the system
\begin{equation}\label{syst}
\begin{cases}
\cl_1\psi=\sigma_1(k)\psi\\
\hbox to 25mm {\dotfill} \relax\\
\cl_n\psi=\sigma_n(k)\psi
\end{cases}
\end{equation}
where $k\in V$ is a "spectral" parameter.
For generic $k$ this space is spanned by the Baker-Akhiezer functions
$\psi(x,g(k)), g \in G$. From the form (\ref{BA}) of this function
it follows that $\psi$ itself must be an eigenvector of $A$:
$$A \psi(x,k) = a(k) \psi(x,k)$$ for some polynomial $a(k)$.
To show that $a(k)$ is a quasiinvariant let us notice
that the left hand side of the last formula satisfies the quasiinvariance conditions in $k$
(see the properties of the BA function)
and therefore must the right hand side. A simple analysis shows
that $a(k)$ must be a quasiinvariant in that case.

Another relation between quasiinvariants and quantum integrals
${\cal L}_q$ is given by the following

\begin{theorem}\label{invsp} The space $Q_m$ of all $m$-quasiinvariants is
invariant under the action of all the operators ${\cal
L}_q$, $q\in Q_m$.
\end{theorem}

\noindent
For the operator ${\cal L}$ (\ref{lcur}) this can be proven by direct
local considerations (c.f. \cite{Ch} where a similar observation
has been first made). The fact that the same is true for any
${\cal L}_q$ now follows from Berest's formula (\ref{chi}).

\section{\pmb{$m$}-harmonic polynomials}

Consider again the space $V(k)$ of the solutions of compatible system of
equations~(\ref{syst}).
Let us put now $k$ to be zero, i.e. consider the system
\begin{equation}\label{syst0}
\begin{cases}
\cl_1\psi=0\\
\hbox to 15mm {\dotfill} \relax\\
\cl_n\psi=0
\end{cases}
\end{equation}
We claim that all the solutions of the system (\ref{syst0}) are
polynomials in $x$. More precisely we have the following

\begin{theorem}  For any Coxeter group $G$ and multiplicity
function $m$ all the solutions of the system (\ref{syst0}) are
polynomial. They form the space of dimension $|G|$
%which as a $G$-module is isomorphic to the group algebra $\C[G]$.
where the natural action of $G$ is its regular representation.
\end{theorem}

When all the multiplicities are zero this is the classical result (see \cite{Steinberg,Helg}) and
the corresponding polynomials are called harmonic.
For the general multiplicity $m$ we will call the corresponding solutions of
(\ref{syst0}) as {\it $m$-harmonic polynomials} and denote the
space $V(0)$ as $H_m$.

To prove the theorem let us consider first the general system (\ref{syst}).
Heckman and Opdam \cite{HO} showed that it is equivalent to a holonomic system of the first
order of rank $|G|$. Components of this system are
$\phi=(\phi_i)$, $\phi_i=q_i(\p)\psi$, where $q_i$, $i=1,\ldots, |G|$ is a
basis of harmonic polynomials of $G$ (see \cite{HO})\footnote{Strictly speaking the authors of \cite{HO}
consider the trigonometric case and Weyl groups but all the arguments work 
for the rational case and any Coxeter group as well. Corresponding holonomic
system can be rewritten in explicit way as a version of Knizhnik--Zamolodchikov equation
(see \cite{FV}).}.

For generic $k$ (more precisely, if
$\prod_\a(k,\a)\ne 0$) we can choose the functions
$\psi_\sigma=\psi(\sigma(k),x), \sigma\in G$, where $\psi$ is the
Baker--Akhiezer function (\ref{BA}) from the previous section, as
a basis of the correspondent space $V(k)$. Since BA
function is regular everywhere as a function of $x$, the same is
true for the solutions of (\ref{syst}) if $\prod_\a(k,\a)\ne 0$.

To prove that this is true for any $k$, in particular for $k=0$,
one can argue as follows.
Let us consider the natural complex version of the system (\ref{syst})
by assuming simply that $x \in V^{\C}$ and $\psi$ takes values in $\C$.
 Consider a point $x_0$ such that
$\prod_\a(x,\a)\ne 0$ and fix the solution of (\ref{syst})
$\psi(k,x; x_0,a), \, a\in\C^{|G|}$ by fixing the initial data in the
corresponding holonomic system $\phi_i(x_0)=a_i$. Since the
system (\ref{syst}) (and the corresponding holonomic system) is
regular in $k$ everywhere  $\psi(k,x; x_0,a)$ is analytic in $k$
everywhere for any $x$ such that $\prod_\a(x,\a)\ne 0$. By Hartogs
theorem (see e.g. \cite{SH}), $\psi(k,x;x_0,a)$ is analytic in $k$
and $x$ everywhere. In particular, $\psi(0,x;x_0,a)$ is analytic
in $x$ at $x=0$. Since the system (\ref{syst0}) is homogeneous,
any component in the Taylor expansion of this function at $x=0$ of
a given degree $d$ is as well a solution of this system. This
proves that all the solutions of (\ref{syst0}) are polynomial.

To prove the second statement of the theorem let us notice that
a natural action of the group $G$ on the space $V(k)$ for a generic $k$
is regular.
This immediately follows from the formula for a basis
$$
\psi_\sigma(x,k) = \psi(x,\sigma(k)), \ \sigma\in G
$$
in terms of BA function.
Indeed, for any $\tau\in G$
$$
\psi_\sigma(\tau^{-1}(x),k) = \psi(\tau^{-1}(x),\sigma(k)) =
\psi(x,(\tau\circ\sigma)(k))=\psi_{\tau \sigma}(x,k)
$$
since $\psi(\tau x,\tau k)=\psi(x,k)$.
For arbitrary $k$ this follows now by
standard continuation arguments.

\section{Quasiinvariants and \pmb{$m$}-harmonic polynomials}

The first relation between the space $H_m$ of $m$-harmonic
polynomials and the space $Q_m$ of $m$-quasiinvariants is given by
the following
\begin{theorem} Any $m$-harmonic polynomial is $m$-quasiinvariant: $H_m\subset Q_m$.
\end{theorem}
We will prove actually the following more general statement.
\\

\noindent
{\bf Proposition 1} {\em Any polynomial $p(x)$ belonging to the kernel
of the operator
$$
{\cal L} = \Delta -\sum_{\a\in A}\frac{2m_\a}{(\a,x)}\p_\a
$$
is a quasiinvariant.}
\\

\noindent
Let us deduce the quasiinvariance condition (\ref{quasi}) for
polynomial $p(x)$ at
 the
hyperplane $(\a,x)=0$. Choose an orthogonal coordinate
system $(t,y_1,\ldots,y_{n-1})$ such that the first axis is normal to the hyperplane.
 Then the operator $\cl$ can be represented as
$$
\cl=\p^2_t+\Delta_y-\left(\frac{2m_\a}{t}+t
f(t^2,y)\right)\p_t+\sum_{i=1}^{n-1} g_i(t^2,y)\p_{y_i},
$$
where $\Delta_y=\p^2_{y_1}+\ldots+\p^2_{y_{n-1}}$. The functions $f$
and $g_i$ are analytic at $t=0$ and invariant under reflection
$t\to -t$ with respect to $(\a,x)=0$ due to invariance of the
operator ${\cal L}$
(c.f. \cite{VSCh}).
For a polynomial $p(x)$ we also have a similar expansion
$$
p=\sum_{i=0}^{\deg p} p_i(y) t^i.
$$
Substituting this into the equation ${\cal L}p=0$ we have
$$
\left(\p^2_t-\frac{2m_\a}{t}\p_t+\Delta_y-t
f(t^2,y)\p_t+\sum_{i=1}^{n-1} g_i(t^2,y)\p_{y_i}\right)
\left(\sum_{i=0}^{\deg p} p_i(y) t^i\right)=0.
$$
Considering all possible terms at $t^{-1}$ in the lefthand side we
conclude that $p_1\equiv 0$. Considering now the terms at $t$
we come to
$$
(6-6m_\a)p_3\equiv 0
$$
which implies that $p_3\equiv 0$ if $m_\a>1$. Continuing in this way we obtain
$$
p_1=p_3=\ldots=p_{2m_\a-1}\equiv 0
$$
or equivalently
$$
\p_\a^{2s-1} p(x)|_{(\a,x)=0}=0  \ \mbox{for} \ 1\le s\le m_\a
$$
which are the quasiinvariance conditions.
Thus the proposition (and therefore the theorem) is proven.

We know the classical fact (see \cite{Steinberg}) that
the space $H_0$ of usual harmonic polynomials related to a Coxeter
group $G$ is isomorphic to the quotient
\beq{factor0}
H_0\approx \C[x_1,\ldots,x_n]/I_0
\eeq
where $I_0$ is the ideal generated by the $G$-invariants of
positive degree. We are going to present some arguments in favour
of the following generalisation of (\ref{factor0}):
\beq{factor}
H_m\approx Q_m/I_m,
\eeq
where $Q_m$ is the ring of all $m$-quasiinvariants, $I_m\subset
Q_m$ is its ideal generated by the invariants of positive degree.

Following the classical scheme \cite{Steinberg} let us define the map
\beq{pim}
\pi_m: Q_m\to H_m
\eeq
by the formula
\beq{pimf}
\pi_m(q) = \cl_q(w_m)
\eeq
where $w_m=\prod_\a (\a,x)^{2m_\a+1}$ and $\cl_q=\chi_m(q)$ is
defined by (\ref{chi}).
To prove that $\cl_q(w_m)\in H_m$ we will need the following
lemma. Let $A_m\subset Q_m$ be the subspace of antiinvariants,
i.e. the quasiinvariants $q$ satisfying the property
$$
q(s_\a(x))=-q(x)
$$
for any reflection $s_\a\in G$.
\\

\noindent
{\bf Lemma 2} {\it $A_m$ is a one-dimensional module over $S^G$
generated by $w_m$.}
\\

\noindent
It is easy to show that such an antiinvariant is divisible by
$w_m$. Since the quotient is $G$-invariant this implies the lemma.
\\

\noindent
{\bf Lemma 3} {\it The quasiinvariant $w_m$ is $m$-harmonic.}
\\

\noindent
Indeed, since all $\cl_i$ are $G$-invariant and preserve the space
$Q_m$ (see theorem \ref{invsp} above) the polynomials $\cl_i(w_m)$ belong to
$A_m$. Since they have degree less than the degree of $w_m$ they
must be zero.
\\

\noindent
{\bf Lemma 4} {\it The space $H_m$ is invariant under the action
of $\cl_q$ for any $q\in Q_m$.}
\\

\noindent
This follows from commutativity of $\cl_q$ and $\cl_i$,
$i=1,\ldots,n$. All this implies
\begin{theorem} The formula
$$
\pi_m(q)=\cl_q(w_m)
$$
defines a linear map from $Q_m$ to $H_m$.
\end{theorem}
Let us discuss the properties of the map $\pi_m$.
\begin{theorem} The kernel of $\pi_m$ contains the ideal
$I_m$.
\end{theorem}
To prove this let us represent any element $q\in I_m$ as
$$
q=\sum_s q_s p_s
$$
where $q_s\in Q_m,\, p_s\in S^G $.
We have
$$
\cl_q(w_m)=\sum_s\cl_{q_s}\cl_{p_s}(w_m)=0
$$
since $\cl_{p_s}(w_m)=0$ due to lemma 3.

Our first two conjectures (see the Introduction) claim that the kernel of $\pi_m$ coincides with
$I_m$ and
that the restriction of $\pi_m$ onto $H_m$:
$$
\pi_m|_{H_m}: H_m\to H_m
$$
is an isomorphism.
This implies that
$$
Q_m/I_m\approx H_m
$$
and that $Q_m$ is generated by $H_m$ as a module over
$S^G$.
Our third conjecture says that this module is actually free (like
in the classical situation $m=0$).

In the next section we are giving the proofs of all our conjectures
for two-dimensional Coxeter groups with constant multiplicity
function.
%$m_\a=m$ for all~$\a$.

\section{Proofs for the dihedral groups}

Consider the dihedral group $G=I_2(N)$ which is a symmetry group of
a regular $N$-gon. It is generated by the plane reflections with
respect to the lines $\alpha_1^j x_1 + \alpha_2^j x_2 =0,\,
\alpha_1^j = -\sin\frac{2\pi j}N, \, \alpha_2^j = \cos\frac{2\pi
j}N, j=0,1,\ldots N-1$. We suppose that the multiplicity function
$m_{\alpha^j}$ is equal to $m\in\Z_{\ge 0}$ for all
$j=0,\ldots,N-1$.
It will be convenient for us to use the complex coordinates
$z = x_1 + i x_2, \bar z = x_1 - i x_2$. The generators of the ring of invariants
can be chosen as $\sigma_1= z \bar z, \sigma_2 = z^N + \bar z^N$.

% (in complex coordninates) $z-e^{\frac{2\pi i}{N}k} \bar z
%0, \, k=0,1,\ldots, N-1$.

%By $m$-quasiinvariants ($m\in Z_+$) we mean the ring $Q$ of
%polynomials $p(x_1, x_2)$ which
%satisfy the equations $\partial_{\alpha_j}^{2s-1} p = 0$ restricted to the
%line $\alpha_1^j x_1 +\alpha_2^j x_2 =0$, $s=1,2,\ldots, m$.

We are going first to investigate the ring $Q_m$ of $m$-quasiinvariants.
For $m=1$ this ring has been studied
in the paper \cite{VK} where in particular the multiplicative generators of $Q_1$ have been found.
We will use some observations from that paper to describe the ring $Q_m$ with general $m$.

Let us consider the following $2(N-1)$ polynomials:
\begin{equation}
\label{qi}
q_j = a_{j0}z^{mN+j} + a_{j1}\bar z^N z^{(m-1)N+j}+\ldots +a_{jm}
\bar z^{mN} z^j,
\end{equation}
\begin{equation}
\label{bqi}
\bar q_j = \bar a_{j0}\bar z^{mN+j} + \bar a_{j1} z^N \bar
z^{(m-1)N+j}+\ldots +\bar a_{jm}
    z^{mN} \bar z^j,
\end{equation}
$j=1,\ldots,N-1$, where the coefficients $a_{js}$ are chosen to
satisfy the system of equations
$$
\left\{\begin{aligned}
&(mN+j)a_{j0}+((m-2)N+j)a_{j1}+ . \, . \, . +(-mN+j)a_{jm}=0\\
&(mN+j)^3 a_{j0}+((m-2)N+j)^3 a_{j1}+  . \, . \, . +(-mN+j)^3 a_{jm}=0\\
&\hbox to 12.5cm {\dotfill}\relax\\
&(mN+j)^{2m-1}a_{j0}+((m-2)N+j)^{2m-1}a_{j1}+  . \, . \, . +(-mN+j)^{2m-1}a_{jm}=0
\end{aligned}
\right.
$$
One can easily see that rank of this system is equal to $m$ so for any $j$ the
coefficients $a_{js}$ are defined uniquely up to proportionality.

Explicitly polynomial $q_j$ can be given as determinant
\begin{equation}
\label{det}
q_j = det Z_j
\end{equation}
of the
following matrix
$$
Z_j=\left(\begin{array}{cccc}
(mN+j)& ((m-2)N+j)&\ldots & (-mN+j)\\
(mN+j)^3& ((m-2)N+j)^3 & \ldots & (-mN+j)^3\\
\vdots&\vdots&\ddots&\vdots\\
(mN+j)^{2m-1}& ((m-2)N+j)^{2m-1} & \ldots & (-mN+j)^{2m-1}\\
z^{mN+j}&\bar z^N z^{(m-1)N+j}&\ldots&\bar z^{mN} z^j
\end{array}
\right).
$$
\\

\noindent
{\bf Proposition 2} {\it The polynomials $q_j, \bar q_j$ belong to the
space $Q_m$ of quasiinvariants.}
\\

\noindent
{\bf Proof.} Let us introduce the polar coordinates $z=re^{i\varphi}$,  $\bar
z=re^{-i\varphi}$. Then from the system of equations defining coefficients
$a_{js}$ it obviously follows that
$\partial_\varphi^{2s-1} q_j |_{\varphi=\frac{\pi k}N} =0$,
$\partial_\varphi^{2s-1} \bar q_j |_{\varphi=\frac{\pi k}N} =0$,
$k=0,\ldots,N-1$, $s=1,\ldots,m$. Now the statement follows from the following
lemma.
\\

\noindent
{\bf Lemma 5} {\it For any polynomial $p(x_1,x_2)$, any vector
$\alpha=(-\sin\varphi_0, \cos\varphi_0)$ and for arbitrary $m\in \Z_+$
the conditions
$$
\partial_\alpha^{2s-1} p|_{(\alpha,x)=0}=0, \quad s=1,\ldots,m
$$
are satisfied if and only if the following conditions in polar
coordinates hold:
$$
\partial_\varphi^{2s-1} p|_{\varphi=\varphi_0}=0, \quad s=1,\ldots,m
$$
}

\noindent
Now let us introduce two more quasiinvariants
\begin{equation}
\label{q0N}
q_0=1,\
q_N=(z^N-\bar z^N)^{2m+1}.
\end{equation}
The last polynomial $q_N$ is actually a basic antiinvariant quasiinvariant
($w_m$ in our previous notations).

\begin{theorem} The ring $Q_m$ is a free finitely generated
module over its subring $S^G\subset Q_m$ of invariant polynomials.
One can choose the polynomials $q_0, q_1,\ldots, q_{N-1}$, \\$\bar
q_1,\ldots, \bar q_{N-1}, q_N$ as a basis of $Q_m$ over $S^G$.
\end{theorem}

\noindent
{\bf Proof.} At first let us show that the polynomials
$q_0, q_1,\ldots, q_{N-1}, \bar
q_1,\ldots, \bar q_{N-1}, q_N$
do generate $Q_m$ over $S^G$.
To prove this we will use induction on a degree of a polynomial
$q\in Q_m$.
If $\deg q=0$ then $q=const=c$, thus $q=c q_0$ so we have checked the
base of induction. Suppose now that $\deg  q =d$ and $q=A z^d+ B\bar
z^d +z\bar z p_{d-2}$
is an arbitrary quasiinavariant, $q\notin S^G$.
We will use further the following two lemmas.
\\

\noindent
{\bf Lemma 6}
{\it Any $m$-quasiinvariant of degree $d\le m N$ is actually
invariant.}
%{$$d>mN$$}
\\

\noindent
{\bf Proof.} It is enough to prove the lemma for an arbitrary
homogeneous polynomial. Let
\begin{equation}
\label{q}
q=a_0 z^{m N-\sigma}+a_1 z^{m N-\sigma-1}\bar z +a_2 z^{m N-\sigma
-2}\bar z^2+\ldots+ a_{m N-\sigma}\bar z^{m N-\sigma} \in Q_m
\end{equation}
for some $\sigma\ge 0$.
According to lemma 5 the conditions of
quasiinvariance in polar coordinates are $\partial_\varphi^{2s-1}q=0$ for
$\varphi = \frac{\pi k}{N}$, $0\le k\le N-1$. We have
$$
(m N-\sigma)^{2s-1}a_0 e^{i\frac{\pi}N(m N-\sigma)k}+
(m N-\sigma-2)^{2s-1}a_1 e^{i\frac{\pi}N(m N-\sigma-2)k}+
$$
$$
(m N-\sigma-4)^{2s-1}a_2 e^{i\frac{\pi}N(m N-\sigma-4)k}+\ldots+
(-m N+\sigma)^{2s-1}a_{m N-\sigma} e^{i\frac{\pi}N(-m N+\sigma)k}
=0
$$
Collecting the terms in this sum with equal exponents, we get
$$
\sum_{j=0}^{N-1}\sum_{\nad{t\ge 0}{j+N t\le m N-\sigma}}(m
N-\sigma-2(j+Nt))^{2s-1}
a_{j+Nt}e^{i\frac{\pi}{N}(mN-\sigma-2(j+Nt))k}=
$$
$$
\sum_{j=0}^{N-1}
e^{i\frac{\pi}{N}(mN-\sigma-2j)k}
\sum_{\nad{t\ge 0}{j+N t\le m N-\sigma}}(m
N-\sigma-2(j+Nt))^{2s-1} a_{j+Nt}=0
$$
Now let us consider these conditions for all possible
$k=0,1,\ldots,N-1$. We arrive at the Vandermonde-type system with
different exponents $e^{i\frac{\pi}N(mN-\sigma-2j)}$, $0\le j\le
N-1$. Hence, for all $j=0,\ldots, N-1$ the following property is
satisfied
\begin{equation}
\label{prop}
\sum_{\nad{t\ge 0}{j+N t\le m N-\sigma}}(m
N-\sigma-2(j+Nt))^{2s-1} a_{j+Nt}=0
\end{equation}
Let us analyze the conditions (\ref{prop}) for all possible $s, \
1\le s \le m$. We again have a system of Vandermonde type with the
exponents $(mN-\sigma-2(j+Nt))^2$, $0\le t\le
[\frac{mN-\sigma-j}N]\le m$. Notice that the exponents
corresponding to different $t$ may coincide only in pairs. The
condition for that is
\begin{equation}
\label{t12}
mN-\sigma-(j+Nt_1)=j+Nt_2.
\end{equation}
In the case $j=\sigma=0$ the number of equations in (\ref{prop})
is less than number of unknown coefficients $a_{j+Nt}$. But after
collecting the terms in (\ref{prop}) corresponding to equal
exponents the number of equations becomes not less than the number
of unknowns. We conclude that all nonzero coefficients $a_{j+Nt}$
can be divided into pairs so that $a_{j+Nt_1}-a_{j+Nt_2}=0$ and
also the condition (\ref{t12}) is satisfied. In terms of
polynomial $q$ this means that it can be represented in the form
$$
q=\sum_{j=0}^{N-1}\sum_{(t_1,t_2)} a_{j+Nt_1}z^{mN-\sigma-(j+Nt_1)}\bar
z^{j+Nt_1}+a_{j+Nt_2}z^{mN-\sigma-(j+Nt_2)}\bar z^{j+Nt_2} =
$$
$$
\sum_{j=0}^{N-1}\sum_{(t_1,t_2)} a_{j+Nt_1}(z\bar z)^{j+Nt_2}
(z^{N(t_1-t_2)}+\bar z^{N(t_1-t_2)}),
$$
where the pairs $(t_1,t_2)$ satisfy (\ref{t12}) and also we
suppose that $t_1>t_2$. Hence polynomial $q$ is an invariant.
\\

\noindent
{\bf Lemma 7} {\it If $q=Az^d+B\bar z^d+z\bar z p_{d-2}$ is $m$-quasiinavariant
of degree $d=mN+l N$,
$1\le l \le m$ then $A=B$.}
\\

\noindent
{\bf Proof.} We will use the notations and scheme of proof of lemma
6. Let us consider $q$ of the form (\ref{q}), now we have
$\sigma=-l N$. As above in lemma 6 the conditions
(\ref{prop}) for $j=0,\ldots,N-1$ should be satisfied. Let us fix
$j=0$, we get
$$
\sum_{t=0}^{m+l}((m+l-2t)N)^{2s-1}a_{Nt}=0
$$
or equivalently
$$
\sum_{t=0}^{[\frac{m+l}2]}
((m+l-2t)N)^{2s-1}(a_{Nt}-a_{N(m+l-t)})=0.
$$
We have got a system of Vandermonde type with different exponents.
For $l \le m$ from that it follows that
$a_{Nt}=a_{N(m+l-t)}$. If $t=0$ we get
$a_0=a_{N(m+l)}$ which completes the proof of the lemma.

To continue the proof of the theorem let us represent $d$ in the form
$d=mN+jN+k$, where $0\le k<N$.
We have to consider few different cases.\\
a) If $k\ne 0$ then $ q-\frac{A}{a_{k0}} q_k(z^N+\bar z ^N)^j -
\frac{B}{\bar a_{k0}}\bar q_k(z^N + \bar z^N)^j = z\bar z
p_1$ for some polynomial $p_1\in Q_m$, here constants $a_{k0}, \bar a_{k0}$ are
defined by the form (\ref{qi}), (\ref{bqi}) of the polynomials $q_k$, $\bar q_k$. Since $\deg  p_1 = d-2<d$
we have done the induction step.\\
b) If $k=0$, $1\le j \le m$ then lemma 7 states that $A=B$, hence
$q-A(z^N+\bar z^N)^{m+j} = z\bar z p_2$, where $p_2\in Q_m$ and it
can be represented as a linear combination of the polynomials
$q_i,\bar q_i$ with invariant coefficients.\\
c) If $k=0$, $j \ge m+1$ then $q-\frac{A-B}2 q_{N} (z^N+\bar
z^N)^{j-m-1}-\frac{A+B}2(z^N+\bar z^N)^{m+j} = z\bar z p_3$,
where $p_3\in Q_m$ and one can apply the induction hypothesis.

Thus we have proved that polynomials $q_i, \bar q_i$ generate $Q_m$ as an
$S^G$--module. Now we are going to show that this module is free.

To see this let us consider arbitrary nontrivial combination of the
polynomials $q_i, \bar q_i$ with invariant coefficients. Since the
 ring of invariants for the dihedral group is the ring freely
generated by two polynomials $\sigma_1=z\bar z$ and
$\sigma_2=z^N+\bar z^N$, the
linear combination takes the form
$$
p_0^1(\sigma_1, \sigma_2) q_0 + p_1^1(\sigma_1,\sigma_2)
q_1+p_1^2(\sigma_1,\sigma_2) \bar q_1+\ldots  +
$$
$$
p_{N-1}^1(\sigma_1,\sigma_2) q_{N-1}+p_{N-1}^2(\sigma_1,\sigma_2)
\bar q_{N-1} +
p_N^1(\sigma_1,\sigma_2) q_N =0,
$$
where for some $s, \epsilon$ we have $p_s^\epsilon \ne 0$.
Also we can suppose that
$p_s^\epsilon$ is not divisible by $\sigma_2$.
Further, let us represent polynomials $p_j$ as combinations of
monomials in $\sigma_1$, $\sigma_2$ and let us move monomials containing
$\sigma_1$ into righthand side. We have then that
$$
r_0^1(\sigma_2) q_0 + r_1^1(\sigma_2)q_1 +r_1^2(\sigma_2) \bar q_1+\ldots
\ldots
+
r_{N-1}^1(\sigma_2)q_{N-1}+r_{N-1}^2(\sigma_2) \bar q_{N-1} +
r_N^1(\sigma_2)q_N
$$
is divisible by $\sigma_1$
and some polynomial $r_s^\epsilon\ne 0$.
Let us consider monomials having degree which is equal to $s$ modulo
$N$. If $1\le s\le N-1$ then $r_s^1(\sigma_2)
q_s+r_s^2(\sigma_2)\bar q_s$
must be divisible by $z\bar z$, which is impossible as
$r_s^1(\sigma_2) q_s$ contains monomial of the form
$\lambda_1 z^{\mu_1}$ and does not contain degrees of $\bar z$, and
$r_s^2(\sigma_2) \bar q_s$ contains monomial of the form
$\lambda_2 \bar z^{\mu_2}$ and does not contain degrees of $z$.
If $s=0$ or $s=N$
then $r_0^1(\sigma_2)+r_N^1(\sigma_2)(z^{(2m+1)N}-\bar
z^{(2m+1)N})$ must be divisible by $z\bar z$, which is possible only
if $r_0^1=r_N^1=0$ but this is not the case.
Thus the theorem is proven.
\\

\noindent
{\bf Corollary}
{\it The Poincare series for the $m$-quasiinvariants of dihedral group $I_2(N)$ is
$$
p(Q_m,t)=\frac{1+2t^{(mN+1)}+\ldots+2t^{(mN+N-1)}+t^{(2m+1)N}}{(1-t^{2})(1-t^{N})}.
$$
}

%Let the polynomials
%$r_0(x), r_N(x)$ have the form $r_0(x)=ax^{n_1}+\mbox{lower
%terms}$, $r_N(x)=bx^{n_2}+\mbox{lower terms}$,

%\bigskip

%\begin{center}
%{\large \bf Quasiinvariants and generalized harmonic polynomials}
%\end{center}

Now we are going to show that polynomials $q_i, \bar q_i$
(\ref{qi}), (\ref{bqi}), (\ref{q0N})
are actually $m$-harmonic. This will complete the proof of
conjecture 3.

First let us rewrite the operator $\cl$ in the complex coordinates.
The set of vectors $\alpha$ for the group $I_2(N)$ has the form
$\alpha = (-\sin\varphi_k, \cos\varphi_k)$, where
$\varphi_k=\frac{\pi k}N$, $k=0,\ldots, N-1$.
Substituting $\partial_x=\partial_z+\partial_{\bar
z}$, $\partial_y=i(\partial_z-\partial_{\bar
z})$ we get
$$
\cl=\Delta - 2m\sum_\alpha\frac{\partial_\alpha}{(\alpha,x)} =
\Delta - 2m\sum_\alpha\frac{-\sin\varphi_k\partial_x+\cos\varphi_k\partial_y}
{-\sin\varphi_k x + \cos \varphi_k y} =
$$
$$
4\partial_z\partial_{\bar z} - 2m\sum_{k=0}^{N-1}
\frac{(-\sin\varphi_k+i\cos\varphi_k)\partial_z+(-i\cos\varphi_k-\sin\varphi_k)\partial_{\bar
z}}
{\frac12 z(-i\cos\varphi_k-\sin\varphi_k)+\frac12\bar
z(i\cos\varphi_k-\sin\varphi_k)}=
$$
$$
4\left(\partial_z\partial_{\bar z} -
m\sum_{k=0}^{N-1}\frac{e^{i\varphi_k}\partial_z-e^{-i\varphi_k}\partial_{\bar
z}}
{-e^{-i\varphi_k}z+e^{i\varphi_k}\bar z}\right).
$$
The operator $\cl=\cl_1$ has a commuting operator ${\cal L}_2$ which is also
invariant under dihedral group, it is homogeneous of degree $N$
and has the form ${\cal L}_2=\partial_z^N + \partial_{\bar z}^N +\mbox{lower
order terms}$.

\begin{theorem} The polynomials $(\ref{qi}), (\ref{bqi}), (\ref{q0N})$ belong to
the common kernel of the operators ${\cal L}_1$ and ${\cal L}_2$, i.e. they are
$m$-harmonic.
\end{theorem}

\noindent
{\bf Proof.}
   From theorem \ref{invsp} and degree consideration it follows immediately that
${\cal L}_1(q_0)={\cal L}_2(q_0)=0$. As
$q_N$ is an antiinvariant quasiinvariant of the smallest possible
degree and due to invariance of the operators ${\cal L}_1, {\cal L}_2$, we have
${\cal L}_1(q_N)={\cal L}_2(q_N)=0$.

Let us show now that
${\cal L}_2(q_s)=0$, $1\le s
\le N-1$. Let ${\cal L}_2(q_s)=r_s, {\cal L}_2(\bar q_s)=\bar r_s$. Notice that
two dimensional space $V_s=<q_s, \bar q_s>$ is an irreducible
representation for the group $G = I_2(N)$. Since operator ${\cal L}_2$ being
invariant commutes with the action of $G$, then by Schur lemma the kernel of
${\cal L}_2|_{V_s}$  is either $V_s$ or $0$. In the last case
the space $<r_s, \bar r_s>$ is an irreducible representation
for $G$. But since $\deg  r_s=\deg  q_s - N< m N$  the polynomials
$r_s, \bar r_s$ should be
invariant according to  lemma 6.
The contradiction means that ${\cal L}_2|_{V_s} = 0$, i.e.
${\cal L}_2(q_s)={\cal L}_2(\bar q_s) =0$.

Now let us show that ${\cal L}_1(q_s)={\cal L}_1(\bar q_s) =0$. Let
$p_s={\cal L}_1(q_s), \bar p_s = {\cal L}_1(\bar q_s)$. As above by
Schur lemma either
${\cal L}_1|_{V_s} = 0$ or  ${\cal L}_1|_{V_s}$ is an isomorphism.
In the second case representation $<p_s,\bar p_s>$ is isomorphic
to irreducible representation $V_s$.
It is easy to see from the formulas (\ref{qi}), (\ref{bqi}) that
among all the representations $V_t, 1\le t\le N-1, t\ne s$ only
$V_{N-s}$ is isomorphic to $V_s$, so that $q_{N-s}$
corresponds to $\bar q_s$, and $\bar q_{N-s}$ corresponds to $q_s$.
%If representation $<p_s, \bar p_s>$ is isomorphic to $V_s$
%then $p_s\to q_s$, $\bar p_s\to \bar q_s$ and since $p_s,
%\bar p_s\in Q$ it's necessary that
This implies that
$p_s=P(\sigma_1,\sigma_2)\bar q_{N-s}$,
$\bar p_s=P(\sigma_1,\sigma_2) q_{N-s}$ for some polynomial
$P(x,y)$.
But since
$$
p_s=
4\left(\partial_z\partial_{\bar z} -
m\sum_{k=0}^{N-1}\frac{e^{i\varphi_k}\partial_z-e^{-i\varphi_k}\partial_{\bar
z}}
{-e^{-i\varphi_k}z+e^{i\varphi_k}\bar z}\right)
(a_{s0}z^{mN+s} + a_{s1}\bar z^N z^{(m-1)N+s}+\ldots +a_{sm}\bar
z^{mN}z^s),
$$
the degree of $p_s$ in $\bar z$ is less or equal then $m N$ while
% $deg_{\bar z}p_s\le mN$, but
$\deg_{\bar z} P(\sigma_1,\sigma_2)\bar q_{N-s}>mN$. This means that
${\cal L}_1|_{V_s}=0$ so  ${\cal L}_1(q_s)={\cal L}_1(\bar q_s)=0$.
The theorem is proven.

The theorems 8 and 9 imply that our conjecture 3 is true for any
dihedral group and constant multiplicity function. Let us now prove the first two
conjectures.

The following lemma will be essential for us.
Let
$$
w = w_m
=\prod_\a(\a,x)^{2m+1}=\prod_{j=0}^{N-1}(-\sin\frac{2\pi j}N
x_1+\cos\frac{2\pi j}N x_2)^{2m+1}
$$
be $m$-discriminant. Let us introduce the corresponding quantum
integral $\cl_w=\chi_m(w)$.
We claim that the constant $\cl_w w$ is non-zero for any $m$.
\\

\noindent
{\bf Lemma 8}
{\em The constant $\cl_w w$ is determined by the formula}
$$
\cl_w(w) = \frac{N^{2m+1}}{2^{(2m+1)(N-1)}} \prod_{j=1}^{2m+1}
(2j-2m-1)\prod_{\nad{d=1}{d\not\equiv 0 (mod N)}}^{(2m+1)N}
(d-m N),
$$
{\em and in particular, is non-zero for any} $m\in\Z_{\ge 0}$.
\\

\noindent
{\bf Remark} When $m=0$ $\cl_w=\prod_\a \p_\a$ and the lemma claims
that
$$
(\prod_\a \p_\a)\prod_\a (\a,x) = \frac{N!}{2^{N-1}}
$$
which is a particular case of the following Macdonald identity:
$$
(\prod_\a \p_\a)\prod_\a (\a,x) = \prod_i d_i!
$$
where $d_i$ are the degrees of generators $\sigma_i$ of the
invariants $S^G$ of a Coxeter group $G$ and all the roots are supposed to be normalised
as $(\a,\a)=2$.
\\

\noindent
{\bf Proof of the lemma.}
Since by formula (\ref{chi}) $\cl_w=\frac1{2^M M!} ad_\cl^{M} w$, $M=\deg w = (2m+1)N$ and
$\cl(w)=0$, we have that
$\cl_w(w)=\frac1{2^M M!}\cl^{M}(w^2)$. Now let us rewrite operator $\cl$ in
polar coordinates. We have
$$
\Delta = \p^2_r+\frac1{r^2}\p^2_\vf+\frac1{r}\p_r.
$$
If $\a=(\cos\vf_0,\sin\vf_0)$ then
$$
\frac{\p_\a}{(\a,x)}= \frac{\cos\vf_0(\cos\vf\p_r-\frac1{r}\sin\vf\p_\vf)+
\sin\vf_0(\sin\vf\p_r+\frac1{r}\cos\vf\p_\vf)}
{r(\cos\vf_0\cos\vf+\sin\vf_0\sin\vf)}=
\frac1{r}\p_r-\frac1{r^2}\tan(\vf-\vf_0)\p_\vf
$$
and
$$
\cl=\Delta-2m\sum\frac{\p_\a}{(\a,x)}=
$$
$$
\p^2_r+\frac1{r^2}\p^2_\vf+\frac1{r}\p_r-\frac{2mN}{r}\p_r-\frac{2mN}{r^2}\cot
N\vf\p_\vf.
$$
Let us notice that $\cl$ maps the space $V^d$ of homogeneous functions
of degree $d$ to $V^{d-2}$, and the restriction of $\cl$ onto $V^d$ takes the
form
$$
\cl|_{V^d}=\frac1{r^2}(d^2-2mNd+\p^2_\vf-2mN\cot N\vf\p_\vf).
$$
Now we are going to calculate  $\cl^M w^2$. In the polar coordinates we have
$$
w^2 = \frac{r^{(2m+1)2N}(\sin N\vf)^{4m+2}}{2^{(N-1)(4m+2)}}
$$
and
$$
\cl^M(w^2)=\frac1{2^{(N-1)(4m+2)}}\left[\prod^{(2m+1)N}_{d=1}
(4d^2-4mNd+\p^2_\vf-2mN\cot (N\vf)\p_\vf)\right]
\sin^{4m+2}N\vf.
$$
Let us introduce new variable $\phi=N\vf$. Also due to
commutativity we may order the terms in the previous product as follows
\begin{multline}\label{zvez}
\cl^M(w^2)=\frac{N^{(4m+2)N}}{2^{(N-1)(4m+2)}}
\prod^{(2m+1)N}_{\nad{d=1}{d\not\equiv 0 (mod N)}}
(4(\frac{d}{N})^2-4m\frac{d}{N}+\p^2_\phi-2m\cot \phi\p_\phi)\times\\
\prod^{2m+1}_{j=1} (4j^2-4mj+\p^2_\phi-2m\cot \phi\p_\phi)
\sin^{4m+2}\phi.
\end{multline}
Now direct calculation shows that
$$
(4j^2-4mj+\p^2_\phi-2m\cot \phi\p_\phi)\sin^{2j}\phi =
2j(2j-2m-1)\sin^{2j-1}\phi.
$$
Thus we can simplify (\ref{zvez}) as
$$
\frac{N^{(4m+2)N}}{2^{(N-1)(4m+2)}}
\prod_{\nad{d=1}{d\not\equiv 0 (mod N)}}^{(2m+1)N}
(4(\frac{d}{N})^2-4m\frac{d}{N}+\p^2_\phi-2m\cot
\phi\p_\phi)c,
$$
where $c=\prod_{j=1}^{2m+1} 2j(2j-2m-1)$.
Finally we get
$$
\cl^M(w^2)=\frac{N^{(4m+2)N}}{2^{(N-1)(4m+2)}}
\prod_{j=1}^{2m+1} 2j(2j-2m-1)
\prod_{\nad{d=1}{d\not\equiv 0 (mod N)}}^{(2m+1)N}
(4(\frac{d}{N})^2-4m\frac{d}{N})=
$$
$$
N^{4m+2}\prod_{j=1}^{2m+1} 2j(2j-2m-1)
\prod_{\nad{d=1}{d\not\equiv 0 (mod N)}}^{(2m+1)N}
d(d-m N) =
N^{4m+2}\prod_{j=1}^{2m+1} 2j(2j-2m-1)
\times
$$
$$
\frac{M!}{N^{2m+1}(2m+1)!}\prod_{\nad{d=1}{d\ne 0 (mod N)}}^{(2m+1)N}
(d-m N) =
M!(2N)^{2m+1}\prod_{j=1}^{2m+1} (2j-2m-1)\prod_{\nad{d=1}{d\not\equiv
0 (mod N)}}^{(2m+1)N}
(d-m N)
$$
Since $\cl_w w = \frac1{2^M M!}\cl^M(w^2)$ we have
$$
\cl_w w =\frac{(2N)^{2m+1}}{2^{(2m+1)N}} \prod_{j=1}^{2m+1}
(2j-2m-1)\prod_{\nad{d=1}{d\not\equiv 0 (mod N)}}^{(2m+1)N}
(d-m N).
$$
This proves the lemma.

Now we are ready to prove Conjecture 1.

\begin{theorem}  For any dihedral group $I_2(N)$
$$
Ker \pi_m = I_m,
$$
where $\pi_m$ is defined by $(\ref{pimf})$ and $I_m$ is the ideal in
$Q_m$ generated by basic invariants $\sigma_1$, $\sigma_2$.
\end{theorem}
{\bf Proof.}
Let us represent an arbitrary quasiinvariant in the form
\beq{qrep}
q=s_0q_0+\sum_{j=1}^{N-1}(s_jq_j+\bar s_j\bar q_j) +s_N q_N
\eeq
where $s_j,\bar s_j$ are invariants and $q_j, \bar q_j$ are
defined by (\ref{qi}), (\ref{bqi}). Suppose that $q\in Ker\pi_m$,
which is $\cl_q w = 0$. As
$$
\cl_q=\cl_{s_0}+\sum_{j=1}^{N-1}(\cl_{s_j}\cl_{q_j}+\cl_{\bar s_j}\cl_{\bar
q_j})+\cl_{s_N}\cl_{q_N},
$$
the condition $\cl_q w=0$ is equivalent to $\cl_{q^H}w=0$, where
$q^H\in H$ is a subsum in (\ref{qrep}) corresponding to those $j$
that $s_j$ (or $\bar s_j$) are constants. Since $q-q^H\in I_m$ it
is sufficient to prove that $\cl_h w\ne 0$ for any $h\in H$.

It is sufficient to consider only the homogeneous $h$. If
$h=const$ then the statement obviously holds. When $h=\mbox{const}\,
w$ it follows from
lemma 8. Suppose now that $h=\lambda_1 q_j+\lambda_2\bar
q_j$ and $\cl_h w=0$.
Let us consider
$$
\cl_{q_{N-j}}\cl_h w = 0 = \cl_{q_{N-j}h} w = \cl_{\lambda_1 q_{N-j}q_j+
\lambda_2 q_{N-j}\bar q_j} w
$$
The formulas (\ref{qi}), (\ref{bqi}) show that
\beq{zv1}
\lambda_1 q_{N-j}q_j+\lambda_2 q_{N-j}\bar q_j = \lambda_1 a_{N-j 0}a_{j0}
z^{(2m+1)N} + (z\bar z)p,
\eeq
where $p$ is some polynomial in $z, \bar z$.
On the other hand, we should have a general representation (\ref{qrep})
for some invariants $s_i, \bar s_i$
\beq{zv}
\lambda_1 q_{N-j}q_j+\lambda_2 q_{N-j}\bar q_j = \sum (s_i q_i
+\bar s_i\bar q_i) + s_0 q_0 +s_N q_N
\eeq
In the last expression
the sum $\sum (s_i q_i +\bar s_i\bar q_i)$
cannot contain monomials $z^{(2m+1)N}, \bar z^{(2m+1)N}$
as $s_i, \bar s_i$
are
nontrivial polynomials of $z\bar z$ which follows from degree
considerations. Suppose that $\lambda_1\ne 0$, then the lefthand
side of (\ref{zv}) contains $z^{2m+1}$ and it does not contain $\bar z^{(2m+1)N}$ (see
(\ref{zv1})). Hence  $s_N$ must be a nonzero constant $c$.
Now
$$
\cl_{q_{N-j}}\cl_h w = \cl_{\sum (s_i q_i+\bar s_i \bar q_i) +s_0}w+
c \cl_{q_N}w = c \cl_{q_N}w
$$
since $\sum (s_i q_i+\bar s_i \bar q_i) +s_0\in I_m$. Due to lemma
8
$\cl_{q_N}w\ne 0$ so $\cl_{q_{N-j}}\cl_h w \ne 0$ which
contradicts the assumption that $\cl_h w =0$. This implies that
$\lambda_1=0$. Similarly multiplying $\cl_h w=0$ by $\cl_{\bar
q_{N-j}}$ we derive that $\lambda_2=0$ which means that $h=0$.
This proves the theorem and conjecture 1 in this case.

Conjecture 2 now simply follows from the previous arguments.
Indeed we have shown in the proof of the previous theorem that if  $\cl_h w=0$
for some $h \in H_m$ then $h=0$. This implies the following

\begin{theorem}
   For any dihedral group the linear map
$$
\pi_m: H_m\to H_m
$$
is an isomorphism.
\end{theorem}

Let us finally show that conjecture $2^*$ also holds.
For that we fix normalisation of basic quasiinvariants as
$$
q_j = z^{mN+j} + z\bar z p_j,
$$
$$
\bar q_j = \bar z^{mN+j} + z\bar z \bar p_j.
$$
Since $q_{j_1}\bar q_{j_2}$ is divisible by $z\bar z$ it is obvious
that $<q_{j_1}, \bar q_{j_2}>=0$.
The consideration of the degrees shows that  $<q_{j_1}, q_{j_2}>=0$ if $j_1+j_2\ne N$. Let us calculate
$<q_{j}, q_{N-j}>$. We have
$$
q_j q_{N-j} = z^{(2m+1)N}
  + (z\bar z)\hat p_j = \frac12 (q_N+\sigma_2^{2m+1}) + (z\bar z)Q_j,
$$
where $\hat p_j$ is some polynomial and $Q_j$ is a quasiinvariant. Hence
$$
<q_j, q_{N-j}> = ({\cal L}_{\frac12 q_N}+{\cal
L}_{\frac12\sigma_2^{2m+1}+ z \bar z Q_j})w =
\frac12{\cal L}_{q_N} w
$$
which is non-zero by lemma 8. This implies the nondegeneracy
of the form $<,>$ and conjecture $2^*$.

\section{Concluding remarks.}

Besides the proof of the conjectures for all Coxeter groups one of the most interesting open
problems is the description of the space of all $m$-harmonic polynomials.

In the classical case $m=0$ this space can be described as the result
of the differential operators with constant coefficients applied to the
discriminant $w=\prod_{\a\in A} (\a,x)$ (see \cite{Steinberg}). In the general case
one can use the operators $\cl_q$ which correspond to the quasiinvariants
but the quasiinvariants themselves
need an effective description.

One of the alternative ideas is to use the relation between the system (\ref{syst}) and
Knizhnik-Zamolodchikov (KZ) equation discovered by Matsuo and Cherednik
(see \cite{M}). The problem with this is that both Matsuo and Cherednik maps
are degenerate when the spectral parameter is zero. Fortunately one can modify the KZ equation
and define a map which is an isomorphism for all values of the spectral parameters (see \cite{FV}).
The possibility to use these modified KZ equations for our problems is now under investigation. 
The latest news in this direction is the calculation
of the Poincare polynomials of $m$-harmonics for all Coxeter groups \cite{FV}.
The answer is written separately for each isotypical component corresponding 
to a given irreducible representation of the Coxeter group.

Another interesting direction is to develop a similar approach for the rings of integrals of other
algebraically integrable quantum problems, in particular for the trigonometric and difference versions of
Calogero-Moser problem related to root systems. In trigonometric case the corresponding
polynomials satisfy certain difference relations which in the rational limit coincide with the
quasiinvariance relations (1) (see \cite{CV},\cite{VSCh}). 
It would be also very interesting to investigate the analogues of harmonic polynomials in relation with 
the ring of quantum integrals for the generalised Calogero-Moser problems related to
the deformed root systems discovered in \cite{VFCh} (see also \cite{CFV}).
 
Finally we would like to mention that for the Weyl groups $G$
the space of usual harmonic polynomials can be interpreted as
cohomology of the generalised flag varieties \cite{BGG}.
It would be interesting to look at the space of
$m$-harmonic polynomials from this point of view.
This is also related to an important question about the multiplication  structure
on $H_m$ induced by the isomorphism with $Q_m/I_m$.

{\bf Acknowledgement.} We are grateful to E. Ferapontov whose question about
geometry of the surfaces parametrised by the solutions of the
system (\ref{syst}) stimulated our investigations.
We have benefited a lot from the numerous discussions
with Yu.Berest and O.Chalykh. We would like to thank
N. Andreev for the computer calculation of $m$-harmonics for root
system $A_2$ which helped us to guess the answer for the dihedral groups.

One of us (APV) is grateful to M.~Shapiro and to G.~Felder for
very useful and stimulating discussions and to Forschungsinstitut f{\"u}r
Mathematik (ETH, Z{\"u}rich) for the hospitality during April 2001
when this paper has been completed.

This work has been partially supported by EPSRC (grant no CR/M69548).

\end{document}